\documentclass[prb,twocolumn,floatfix,showpacs]{revtex4-1}
\usepackage{graphicx,color}
\usepackage{amsmath,amssymb,bm}

\newcommand{\tn}{\textnormal}
\newcommand{\cprb}[3]{Phys.~Rev.~B {\bf #1}, #2 (#3)}
\newcommand{\cprl}[3]{Phys.~Rev.~Lett.~{\bf #1}, #2 (#3)}

\newcommand{\cjpa}[3]{J.~Phys.~A {\bf #1}, #2 (#3)}
\newcommand{\cepj}[3]{Eur.~Phys.~J.~Spec.~Topics {\bf #1}, #2 (#3)}
\newcommand{\cjpamg}[3]{J.~Phys.~A: Math.~Gen {\bf #1}, #2 (#3)}
\newcommand{\cpre}[3]{Phys.~Rev.~E {\bf #1}, #2 (#3)}

\definecolor{darkred}{rgb}{0.90,0,0}
\definecolor{darkgreen}{rgb}{0,0.60,.2}
\definecolor{darkblue}{rgb}{0,0,1}
\definecolor{grey}{cmyk}{0,0,0,0.25}
\definecolor{orange}{cmyk}{0,0.6,0.8,0}

\begin{document}
\title{\boldmath Nonequilibrium thermal transport and its relation to linear response}
\author{C.\ Karrasch$^{1,2}$}
\author{R.\ Ilan$^{1}$}
\author{J.\ E.\ Moore$^{1,2}$}

\affiliation{$^1$Department of Physics, University of California, Berkeley, California 95720, USA}

\affiliation{$^2$Materials Sciences Division, Lawrence Berkeley National Laboratory, Berkeley, CA 94720, USA}

\begin{abstract}

We study the real-time dynamics of spin chains driven out of thermal equilibrium by an initial temperature gradient $T_L\neq T_R$ using density matrix renormalization group methods. We demonstrate that the nonequilibrium energy current saturates fast to a finite value if the linear-response thermal conductivity is infinite, i.e.~if the Drude weight $D$ is nonzero. Our data suggests that a nonintegrable dimerized chain might support such dissipationless transport ($D>0$). We show that the steady-state value $J_E$ of the current for arbitrary $T_L\neq T_R$ is of the functional form $J_E=f(T_L)-f(T_R)$, i.e.~it is completely determined by the linear conductance. We argue for this functional form, which is essentially a Stefan-Boltzmann law in this integrable model; for the XXX ferromagnet, $f$ can be computed via thermodynamic Bethe ansatz in good agreement with the numerics. Inhomogeneous systems exhibiting different bulk parameters as well as Luttinger liquid boundary physics induced by single impurities are discussed briefly.

\end{abstract}

\maketitle


\section{Introduction} One-dimensional (1d) electronic systems are realized in carbon nanotubes and individual polymer molecules and provide an approximate description of strongly anisotropic 3d materials. It has been known for many years that 1d systems can support unusual correlated electron phenomena such as Luttinger liquid physics. However, electrical and thermal transport in real materials are usually not governed by the free low-energy Luttinger liquid fixed point but by an interplay between dangerously irrelevant operators scattering the currents and conserved quantities protecting them.\cite{andrei,sirker,prosennonloc}

In order to connect to actual experiments, it is thus essential to study generic microscopic models. Over the last decades a significant number of works \cite{bethespin,qmcspingros,qmcspinsorella,edspinmillis,fabianprb,fabianrev,rosch,sirker,prosennonloc,rigol,drudepaper} investigated equilibrium charge (or spin) transport properties. In particular, the question whether or not so-called integrable models, which possess a complete set of local conserved quantities, can support dissipationless currents at finite temperature was addressed extensively. Less is known about the quantitative effects of integrability-breaking perturbations which are naturally present in any experimental system, and even the qualitative question whether the linear-response conductivity of a nonintegrable model can still be infinite is not answered conclusively.\cite{integrability} While experimental measurements of thermal transport driven by a temperature gradient in quasi-1d spin systems already exist,\cite{ott,ott2,hess,hess2} only a few works investigate this theoretically.\cite{fabianprb,fabianrev,thermconserved,bethetherm,qmctherm,roschtherm,orignac,saito,chernyshyev,rosch3} Studying nonequilibrium thermal (or charge) transport is complicated in general -- one reason being that is not even clear whether the long-time dynamics can be described by a low-energy theory -- and constitutes one of the most active areas of research in strongly correlated condensed matter physics.\cite{fabiannoneq,fabiannoneq2,noneqtherm1,noneqtherm2,schmitteckert,noneqprosen1,noneqstein,prosentherm,free2,bruneau,doyon,integrabilitypaper}

The primary goal of our work is to obtain quantitative results on steady-state energy flow both near and far from equilibrium and to understand the effects of integrability and correlations. This is motivated by the experiments listed above and by recent technical advances in dynamical simulations.\cite{drudepaper} As prototypical models we consider a XXZ spin-$1/2$ chain in the presence of two perturbations (dimerization and a staggered magnetic field) which break integrability\cite{levelstat,integrabilitypaper} as well the quantum Ising model. Using density matrix renormalization group methods we demonstrate that the nonequilibrium energy current driven by a temperature gradient $T_L\neq T_R$ relaxes fast to a finite steady-state value if the linear-response thermal conductivity is infinite,\cite{fabiannoneq2} i.e.~if the Drude weight $D$ is nonzero. Our data indicate that the dimerized chain might support such dissipationless transport ($D>0$) despite the fact that it is nonintegrable ($D$ can be extracted from the asymptote of the equilibrium energy current correlation function,\cite{fabianprb} and we cannot exclude that the latter decays on a hidden large temperature-independent time scale).

One of our main results is that for a large class of problems the steady-state current takes, within numerical accuracy, the functional form
\begin{equation}
J_E(T_L,T_R) = f(T_L) - f(T_R)~.
\label{ffunction}
\end{equation}
In words, its dependence on the two temperatures is tightly constrained: The steady-state current is the difference between the total radiated power from the left and right leads.  The function $f$ is thus a generalization of the Stefan-Boltzmann law for photons, for which $f \sim T^{d+1}$ in $d$ spatial dimensions. Moreover, Eq.~(\ref{ffunction}) implies that nonequilibrium thermal transport is entirely determined by linear response -- $f$ can simply be obtained by integration of the equilibrium conductance $\partial_T f$. 

We give an intuitive argument for the existence of a Stefan-Boltzmann function $f$ and also shows that for the XXX ferromagnet, $f$ can be estimated via thermodynamic Bethe ansatz in good agreement with the numerics at low temperatures.  We demonstrate that at low temperatures the gapless integrable XXZ chain as well as the quantum Ising model exhibit universal nonequilibrium behavior conjectured by conformal field theory,\cite{sotiriadis,cardyprize,doyon} which provides a check on the accuracy of the numerical calculations. We finally study inhomogeneous systems featuring different bulk interactions as well as the long-studied Luttinger liquid physics\cite{kanefisher} induced by an impurity at the interface.

\section{Thermal non-equilibrium setup} We aim at investigating the real-time dynamics of the energy current $\langle J_E(n,t)\rangle$ through a one-dimensional infinite lattice system driven out of equilibrium by an initial sharp temperature gradient $T_L\neq T_R$. Our main focus is to study the long-time behavior of $\langle J_E(n,t)\rangle$ and specifically the question how it relates to linear-response thermal transport properties. As a prototypical model, we consider a chain of interacting spin-$1/2$ degrees of freedom $S^{x,y,z}_n$ governed by local Hamiltonians
\begin{equation}\label{eq:h}
h_{n}  =  J_n \big(S^x_{n}S^x_{n+1} + S^y_{n}S^y_{n+1} + \Delta_n S^z_{n}S^z_{n+1}\big) + b_n (S_n^z-S_{n+1}^z) ~,
\end{equation}
or equivalently spinless Fermions through a Jordan-Wigner transformation. By choosing the couplings $J_n$, $\Delta_n$, and $b_n$ appropriately:
\begin{equation}\label{eq:parahom}
J_n =
\begin{cases}
1 & n \tn{ odd} \\
\lambda & n \tn{ even}    
\end{cases}~,~~
\Delta_n = \Delta~,~~ b_n= \frac{(-1)^n b}{2}~~,
\end{equation}
we can study systems which are gapless or gapped and -- as a key aspect of this work -- investigate the role of integrability. For $\lambda=1$ and $b=0$, Eq.~(\ref{eq:h}) can be diagonalized via Bethe ansatz;\cite{bethegs} the model is nonintegrable otherwise. The spectrum is gapless for $|\Delta|\leq1$ and gapped for $\Delta>1$. A gap opens for $\lambda<\lambda_c$ or $b>b_c$, where $\lambda_c<1$ and $b_c>0$ only if $-1<\Delta<-1/\sqrt{2}$.\cite{sato,stagfield,integrabilitypaper} In addition, we study the quantum Ising model
\begin{equation}
h_n = - 4 S^z_{n}S^z_{n+1} -  ( S_n^x+S_{n+1}^x )~.
\end{equation}

Thermal nonequilibrium is introduced via the following protocol: We initially consider two seperate semi-infinite chains ($N\to\infty$)
\begin{equation}
H_0 = H_L + H_R = \sum_{n=-N/2+1}^{-1} h_n + \sum_{n=1}^{N/2-1} h_n ~, 
\end{equation}
each being in thermal (grand-canonical) equilibrium at temperatures $T_L$ and $T_R$. The corresponding density matrix factorizes,
\begin{equation}
\rho_0=\rho_L\otimes\rho_R~,~~\rho_i=\frac{\exp(-H_i/T_i)}{\tn{Tr}\exp(-H_i/T_i)}~. 
\end{equation}
At time $t=0$, the chains are coupled through $h_0$, and the time evolution of $\rho_0$ is computed w.r.t.~$H=H_0+h_0$. The energy current is defined by a continuity equation,\cite{fabianprb}
\begin{equation}
\partial_t h_n = J_E(n) - J_E(n+1) ~\Rightarrow ~ J_E(n) = i [h_{n-1}, h_n]~,
\end{equation}
and its time evolution is simply given by
\begin{equation}
\langle J_E(n,t) \rangle = \tn{Tr} \left[ e^{iHt} \rho_0 e^{-iHt} J_E(n) \right]~,
\end{equation}
which can be computed efficiently using the real-time \cite{tdmrg} finite-temperature \cite{dmrgT} density matrix renormalization group \cite{white,dmrgrev} (DMRG) algorithm introduced in Ref.~\onlinecite{drudepaper}. DMRG is essentially controlled by the so-called discarded weight $\epsilon$. We ensure that $\epsilon$ is chosen small enough and that $N$ is chosen large enough to obtain numerically-exact results (i.e., $\langle J_E(t)\rangle$ to an accuracy of one percent) in the thermodynamic limit. We stop our simulation once the DMRG `block Hilbert space dimension' has reached values of about 1000.

\begin{figure}[t]
\includegraphics[width=0.95\linewidth,clip]{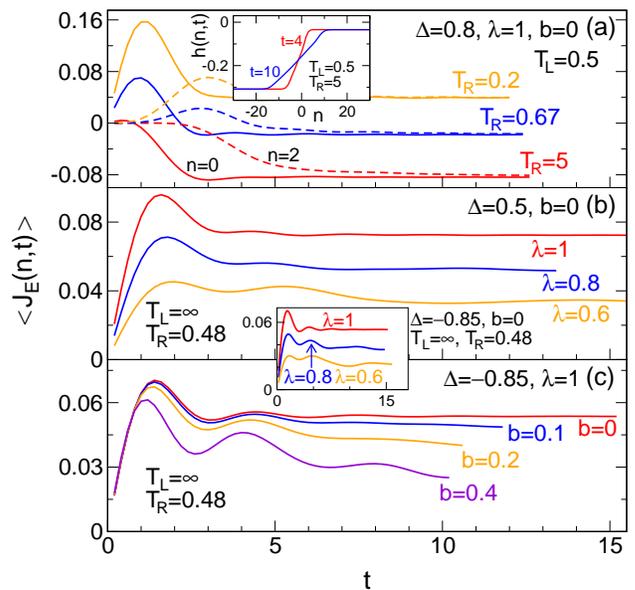}
\caption{Energy current flowing between two semi-infinite spin-$1/2$ chains which are initially in thermal equilibrium at different temperatures $T_{L,R}$ and coupled at time $t=0$ and position $n=0$. (a) Integrable XXZ chain with $z$-anisotropy $\Delta$. The behavior in the gapped phase $\Delta>1$ is similar. (b, lower Inset) Nonintegrable dimerized XXZ chain where the coupling on every second bond is reduced by $\lambda$. The latter is irrelevant at $\Delta=-0.85$ but opens a gap for $\Delta=0.5$. (c) XXZ chain in presence of a staggered field $b$ rendering the model nonintegrable. A gap opens around $b\approx 0.3$. Despite the fact that the local energy density $h(n,t)$ does not relax (upper Inset), the current saturates fast to a unique finite value except for $b>0$. We attribute this to a finite linear-response thermal Drude weight of both the pure XXZ chain and the nonintegrable dimerized chain (see Figure \ref{fig:drude}). }
\label{fig:current}
\end{figure}

\section{Non-equilibrium energy current} We start by studying a XXZ chain with two additional perturbations (dimerization $\lambda<1$ and a staggered field $b>0$) which both render the system nonintegrable.\cite{fabianprb,levelstat,integrabilitypaper} At time $t=0$, two semi-infinite chains each being prepared in thermal equilibrium at temperatures $T_{L,R}$ are coupled by $h_0$ to an overall translationally-invariant chain. Exemplary results for $\langle J_E(n,t)\rangle$ are shown in Figure \ref{fig:current}. The current at the interface $n=0$ saturates on a scale $t\sim1$ [note the definition of units via Eq.~(\ref{eq:parahom})] irrespective of the temperature difference $T_L-T_R$ or the absolute values of $T_{L,R}$ and regardless of the fact whether or not the system is gapped. The only exception is $b>0$ where $\langle J_E(n,t)\rangle$ does not reach a finite steady-state value within the time scales accessible by our numerics [Figure \ref{fig:current}(c)], again irrespective of the fact whether or not $b$ opens a gap. We will now try to understand this in more detail.
 
\begin{figure*}[t]
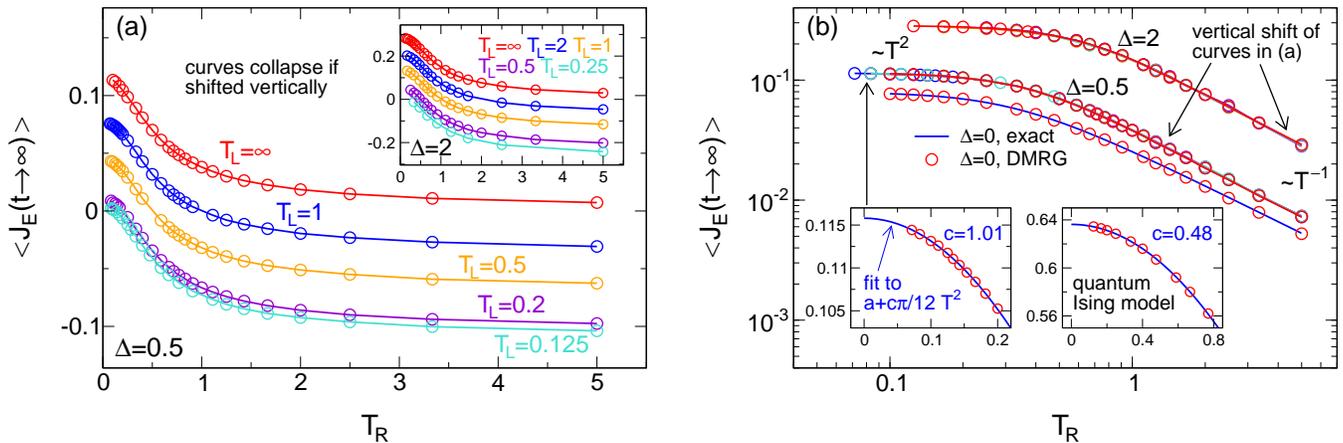

\includegraphics[width=0.475\linewidth,clip]{steady1.eps} \hspace*{0.025\linewidth}
\includegraphics[width=0.475\linewidth,clip]{steady2.eps}
\caption{(a) Temperature-dependence of the steady-state energy current (which becomes position-independent for $t\to\infty$) of the integrable XXZ chain in the gapless ($\Delta=0.5$) and gapped ($\Delta=2$) regime. The right temperature $T_R$ is varied at fixed $T_L$. (b) The curves at different $T_L$ collapse if shifted vertically; thus, the $T$-dependence is of the simple functional form $\lim_{t\to\infty}\langle J_E(n,t)\rangle = f(T_L)-f(T_R)$. The linear thermal conductance $\sim\partial_Tf$ therefore completely determines the nonequilibrium current. A CFT approach \cite{doyon} predicts universal behavior $f(T)=c\frac{\pi}{12} T^2$ at low $T$ which we verify (Insets) for the gapless XXZ chain (central charge $c=1$) and the quantum Ising model ($c=1/2$). The free fermion case $\Delta=0$ can be solved analytically \cite{free2,bruneau} at any $T$ and provides a test for our DMRG numerics. }
\label{fig:steady}
\end{figure*}

The time evolution of the local energy density $h(n,t) = \langle h_n(t)\rangle$ of the XXZ chain (which for a homogeneous system might be a measure for an effective temperature) is shown in the Inset to Figure \ref{fig:current}(a). It does not reach a steady-state value but becomes increasingly smooth. This is not suprising since we are simulating a closed quantum system -- but gives rise to the questions: (1) Why does the current saturate except for $b>0$, implying that it is \textit{not determined by local temperature gradients}? (2) Would we obtain the same steady-state current if we kept the `reservoirs' at a fixed temperature?\cite{prosentherm} Both are reasonable if it does not matter over which length scale $L$ the temperature difference $T_L-T_R$ is applied; qualitatively, this should be the case if thermal transport properties of the chain are length-independent, i.e.~if the thermal conductance $G$ of a finite system does not decrease with its length $L$, or equivalently, if the conductivity $\sigma=GL$ of an infinite chain $L\to\infty$ is infinite. More quantitatively, we conjecture a relation between nonequilibrium and linear response: The nonequilibrium energy current relaxes to a finite steady-state value if the linear-response thermal conductivity is infinite, i.e.~if the Drude weight $D$ is nonzero. Before we proceed with calculating $D$, we note that (2) can be shown explictly for the XX chain $\Delta=b=0$, $\lambda=1$ by carrying out the so-called wide-band limit (which strictly pins the temperatures) and by computing the current analytically using Keldysh Green functions; moreover, the nonequilibrium steady state current was recently obtained from a generalized Landauer-Buettiker formula.\cite{bruneau} Both currents agree with the one in our setup at long times [see, e.g., Figure \ref{fig:steady}(b)].

\begin{figure}[t]
\includegraphics[width=0.95\linewidth,clip]{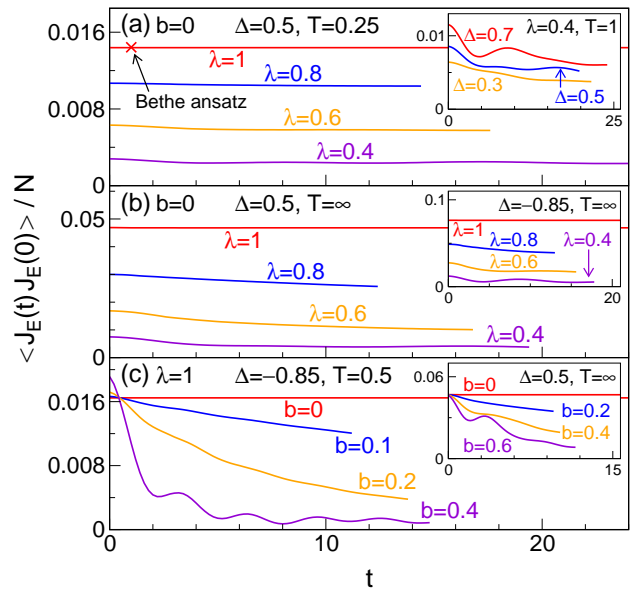}
\caption{Linear-response energy current correlation function whose long-time asymptote determines the Drude weight through Eq.~(\ref{eq:ecorr}). For the integrable XXZ chain ($\lambda=1$, $b=0$), the global energy current $J_E=\sum_n J_E(n)$ is conserved; thus, $\langle J_E(t)J_E\rangle=\langle J_E(0)J_E\rangle$, and DMRG can be compared with exact Bethe ansatz results (symbol). (a,b) Nonintegrable dimerized chain. The current correlation function seems to saturate at a finite value (or decays on a hidden large temperature-independent time scale), indicating a finite Drude weight $D>0$. (c) The data in presence of a staggered field $b>0$ is consistent with $D=0$.}
\label{fig:drude}
\end{figure}

\section{Linear response thermal Drude weight} To support our conjecture we now extract $D$ from the long-time behavior of the energy current correlation function,\cite{sirker,fabianprb}
\begin{equation}\label{eq:ecorr}
D = \lim_{t\to\infty}\lim_{N\to\infty} \frac {\tn{Re } \langle J_E(t)J_E\rangle}{2NT^2}~,~J_E = \sum_n J_E(n)~,
\end{equation}
which can be readily computed using DMRG. Results are shown in Figure \ref{fig:drude}. For $\lambda=1$ and $b=0$, $J_E$ is conserved, thus $\langle J_E(t)J_E\rangle=\langle J_E(0)J_E\rangle$; the Drude weight can alternatively be obtained via Bethe ansatz.\cite{bethetherm} The energy current correlation functions of the dimerized chain seem to saturate at a finite value for any $T$; alternatively, they decay on a hidden large time scale which is temperature-independent and becomes \textit{larger} as the dimerization is increased from $\lambda=1$ to $\lambda=0.4$ [see Figure \ref{fig:drude}(a) and (b)]. Our data thus indicate a nonzero Drude weight. This is interesting on general grounds because the model is nonintegrable.\cite{integrability} Most previous numerical works on the dimerized chain \cite{fabianprb,fabianrev} yield $D=0$ \cite{qmctherm} but focus on $\Delta=1$ where also our results are less conclusive. The deeper reason for a potentially finite Drude weight -- the protection of the energy current by an unknown nonlocal conserved operator \cite{prosennonloc} -- will be left as a subject for future work. In contrast, our data for $b>0$ is consistent with $D=0$ [see Figure \ref{fig:drude}(c)].

Note that in both cases it does not seem to play a role whether $\lambda<1$ and $b>0$ are irrelevant [main panel of Figure \ref{fig:drude}(c); Inset to (b)] or open a gap [main panels of (a) and (b); Insets to (a) and (c)]: If $T$ is decreased in a regime where $b$ is relevant [e.g.~at $\Delta=0.5$; see the Inset to Figure \ref{fig:drude}(c) for $T=\infty$], the scale on which $\langle J_E(t)J_E\rangle$ decays becomes \textit{successively} larger; for temperatures smaller than the gap it can no longer be reached by our numerics. The behavior of the dimerized chain for parameters where $\lambda$ is relevant is completely different: if $\langle J_E(t)J_E\rangle$ decays on a hidden large scale, the latter is temperature-independent and does not manifest even at $T=\infty$ [compare Figure \ref{fig:drude}(a) and (b)].

Recalling that the nonequilibrium current relaxes to a nonzero steady-state value in all cases where $b=0$, the observation of a finite (vanishing) Drude weight for $\lambda\leq1$ ($b>0$) supports our above conjecture.

\section{Asymptotic current, homogeneous system} 
\label{sec:homogeneous}

\subsection{Numerical Results}
We now turn to study the temperature-dependence of the steady-state (position-independent) current. The result for the XXZ chain both in the gapless and gapped regime is illustrated in Figure \ref{fig:steady}. The asymptotic current seems to be of a strikingly simple functional form:
\begin{equation}\label{eq:f}
 \lim_{t\to\infty} \langle J_E(n,t)\rangle = f(T_L) - f (T_R)~,
\end{equation}
indicating a second relation between nonequilibrium and linear response: The linear thermal conductance $\sim \partial_T f(T)$ determines the steady-state nonequilibrium current at any $T_L-T_R$. Equation (\ref{eq:f}) can be established by varying $T_R$ at fixed $T_L$; the corresponding curves collapse if shifted vertically [see Figure \ref{fig:steady}(b)]. The limiting behavior (both in the gapless and gapped regime) of $f(T)$ is given by
\begin{equation}\label{eq:fasym}
f(T) \sim \begin{cases} T^2 & T \ll 1 \\ T^{-1} & T \gg 1~. \end{cases}
\end{equation} 
Other details of $f$ such as prefactors or the crossover scale (which for $\Delta=0$ and $\Delta\gg1$ is determined by the bandwidth or the size of the gap, respectively) in general depend on the model parameters. However, a recent conformal field theory approach\cite{doyon} conjectures that the low-temperature behavior of a gapless system is universally given by $f(T) = c \frac{\pi}{12}T^2$ with $c$ being the CFT central charge, which follows intuitively from the version of the Stefan-Boltzmann law satisfied by a CFT.\cite{sotiriadis,cardyprize} We confirm this prediction for the XXZ chain ($c=1$) as well as the quantum Ising model ($c=1/2$); this is illustrated in the Insets to Figure \ref{fig:steady}(b). This is a nontrivial result because: (1) It is unclear why for a microscopic model whose equilibrium physics is governed by a certain low-energy field theory the very same field theory should describe the long-time behavior of the microscopic model in nonequilibrium (note that the behavior for $0<b<b_c$ is not captured by the CFT!), and (2) Even linear-response transport properties (such as the Drude weight) are not determined by the low-energy theory alone but by a delicate interplay of conserved quantities protecting the current and dangerously irrelevant operators scattering it.\cite{andrei,sirker,prosennonloc}

Our results for the dimerized chain at $\lambda=0.8$ are still consistent with Eq.~(\ref{eq:f}), indicating that it might be a universal property of any system with a thermal Drude weight, if indeed that system has a Drude weight.  At smaller $\lambda$ and low $T$, we cannot reach time scales where oscillations of the current have died out completely.  We expect that models which are strongly nonintegrable and have zero Drude weight will not show a steady state even in the homogeneous case.~\cite{integrabilitypaper} 

The free fermion case $\Delta=0$ can be solved exactly;\cite{free2,bruneau} Eq.~(\ref{eq:f}) reflects a noninteracting thermal Landauer-B\"uttiker formula. This analytic result can be used to test our DMRG numerics at any temperature [see the comparison in Figure \ref{fig:steady}(b) as well as in the Inset to Figure \ref{fig:inhom}].

\begin{figure}[t]
\includegraphics[width=0.7\linewidth,clip]{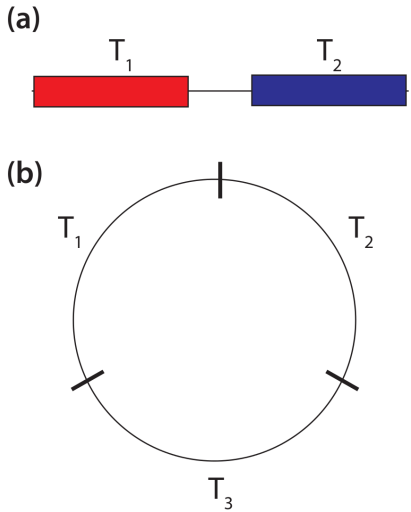}
\caption{ (a) The setup studied in this paper: at $t=0$ two (effectively semi-infinite) reservoirs featuring different temperatures $T_L$ and $T_R$ are connected, and eventually a steady-state energy current is established. (b) Gedanken experiment with three reservoirs.  As described in the text, conservation of the energy current implies that there is a relationship between the currents flowing at the three interfaces, until such times as the different interfaces begin to interact with each other. }
\label{fig:geometry}
\end{figure}

\begin{figure}[b]
\includegraphics[width=0.95\linewidth,clip]{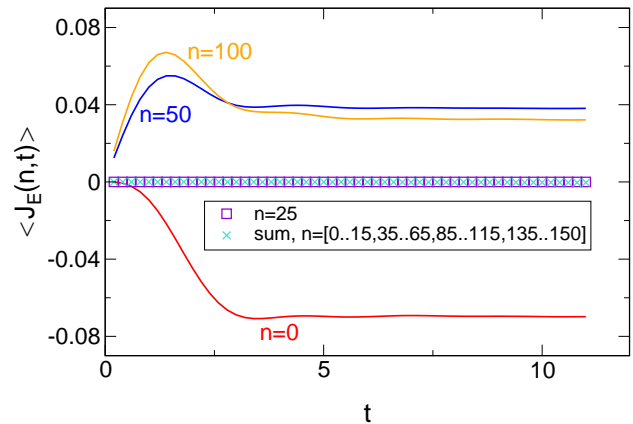}
\caption{ DMRG data for the setup depicted in Figure \ref{fig:geometry}(b). At $t=0$, three XXZ chains ($\Delta=0.5$) of lengths $N=50$ and temperatures $T_1=\infty$, $T_2=1$, $T_3=0.5$ are coupled to a ring. The energy currents saturate to their unique steady-state values at the interfaces ($n=0,50,100$) but remain zero far away from them (e.g., for $n=25$). The sum of the current over the regions where it is nonzero vanishes [see Eq.~\ref{eq:currentsum}] due to the conservation of the total current. As explained in the main text, this motivates the ``cyclic sum rule'' of Eq.~(\ref{eq:cyclicform}), which is equivalent to the existence of a Stefan-Boltzmann function $f$.}
\label{fig:ring}
\end{figure}

\subsection{Stefan-Boltzmann function in integrable systems}

Eq.~(\ref{eq:f}) can alternately be understood as a cyclic relation for the 3-reservoir geometry in Figure~\ref{fig:geometry}(b): The existence of a Stefan-Boltzmann function $f$ is equivalent to the statement that the steady-state currents between three reservoirs $T_{1,2,3}$ satisfy
\begin{equation}
J_E(T_1 \rightarrow T_2) + J_E(T_2 \rightarrow T_3) + J_E (T_3 \rightarrow T_1) = 0~.
\label{eq:cyclicform}
\end{equation}
We now show that integrable models with a conserved total energy current, such as the XXZ model, have a cyclic ``sum rule'' structure which is loosely similar to but not (at first glance) equivalent to Eq.~(\ref{eq:cyclicform}). Further physically motivated assumptions then lead to the existence of the Stefan-Boltzmann function.

To understand why there is any relationship between the three pairs of temperatures in the cyclic formula Eq.~(\ref{eq:cyclicform}), consider the initial condition shown in Fig.~\ref{fig:geometry}(b).  Three segments of equal lengths $N$ of a ring are prepared at three different temperatures. This system can be studied straightforwardly via DMRG, and examplary results are shown in Figure \ref{fig:ring}. They can be interpreted as follows. Suppose that the typical velocity of the system is $v_\tn{typ}$, and consider energy currents at a time $v_\tn{typ} t \ll N$.  For any such time, the middle of each reservoir is essentially unperturbed from its initial state, so the local energy current is zero (see the curve at $n=25$).  The local energy currents rise in the vicinity of the junctions ($n=0,50,100$); let $j_t^\Sigma(T_1 \rightarrow T_2)$ be the {\it spatially integrated} energy current at the $T_1$ to $T_2$ boundary at time $t$, and similarly for the other boundaries.  Of course once $v_{\rm typ} t \approx N$ the energy currents are no longer localized in the region of the boundaries and $j_t^\Sigma$ is not defined.

The initial energy current around the ring is zero and is a conserved quantity of the connected system, so the total energy current must still be zero.  In other words, for all times $v_{\rm typ} t \ll N$ we have (see Figure \ref{fig:ring})
\begin{equation}
j^\Sigma_t(T_1 \rightarrow T_2) + j^\Sigma_t(T_2 \rightarrow T_3) + j^\Sigma_t (T_3 \rightarrow T_1) = 0~.
\label{eq:currentsum}
\end{equation}
This is not the same as Eq.~(\ref{eq:cyclicform}), however, because $J_E$ is the current at a point, while $j^\Sigma_t$ is spatially integrated and has some complicated profile in general.  For the conformal field theory limit, all excitations have a single velocity $v$ and $j^\Sigma_t$ is expected to be dominated by $v t J_E$.  To summarize, general principles suggest that there is a sum rule (at all times less than an upper cutoff determined by the reservoir size) relating the three pairs of temperatures; however, the detailed form of this sum rule is not the same as the observed relationship between steady-state currents, although they are related in the conformal limit.

Now we proceed to give a definition of the Stefan-Boltzmann function $f$ and then argue that the steady-state current is determined by the difference in the Stefan-Boltzmann functions of the two reservoirs.  Consider a large reservoir of size $L_0$ prepared at initial temperature $T$.  At $t=0$, it is connected to semi-infinite leads at each end, and these leads are prepared at some reference temperature, say 0 (ignoring any subtleties from possible symmetry breaking).  At some very long time so that no excitations remain in the reservoir, let $F$ be the total energy current integrated over all sites to the right of the reservoir.  This ``right-moving energy current'' will be extensive in $L_0$, so we define $f = F / L_0$ as the radiated right-moving energy current per site.

In order to give some intuition for the existence of a steady state and Stefan-Boltzmann law even when there is not a single velocity as in the conformal case, we present a free-particle example in Appendix A.  This model has a Gaussian distribution of velocities and a steady-state energy current at the reservoir boundary that is determined solely by the total radiated power as described above.  The variation in particle velocity does not otherwise affect the answer. The steady state lives for a time that is arbitrarily long as the reservoir size becomes infinite, but is short compared to the amount of time it takes for the reservoir to radiate all its energy, as at that time there is no current left at the reservoir boundaries.  This notion of the steady state as actually a long-time phenomenon compared to transients (and diverging in the limit that the reservoir is infinite), but a short-time phenomenon compared to the time on which $F$ is defined, is used in our argument for the interacting case below.

\begin{figure}[t]
\includegraphics[width=0.95\linewidth]{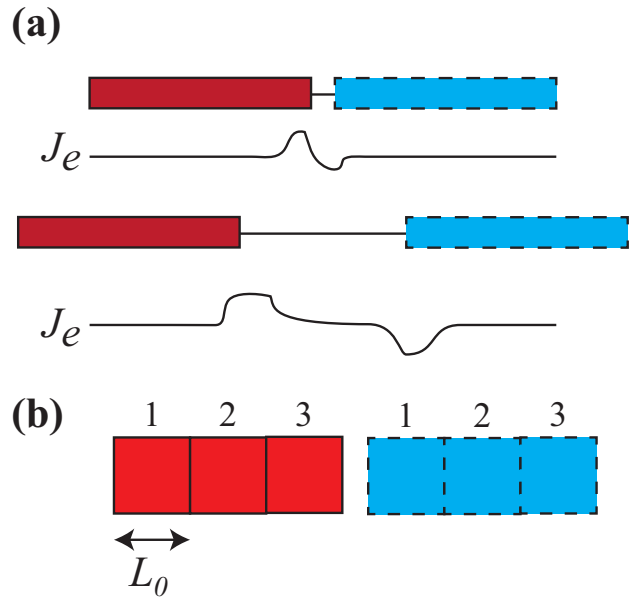}
\caption{(a) Robustness of total integrated current around a junction: the spatial distribution (drawn schematically for illustrative purposes) of the energy current $J_E$ is different between the geometries with and without a ``spacer'' region prepared at some reference temperature, but the spatial integrals are the same. (b) Two large reservoirs of size $L$ can be viewed as made up of translated pairs of smaller sub-reservoirs of size $L_0$.  The numbers indicate the pairings of the sub-reservoirs.  The sum of many translated copies of a function approaches a constant, as long as the function is smoothly varying on the scale of the translations.  If the energy current can indeed be viewed as such a sum, then the existence of a Stefan-Boltzmann function is recovered, independent of the precise shape of the energy current distribution.}
\label{fig:spacer} 
\end{figure}

Now comes a subtle and surprising point.  The cyclic relation at finite times in (\ref{eq:currentsum}) means that the behavior of the energy current at any of the junctions is remarkably constrained: any change that only affects one junction but not the other two will not affect the spatially integrated current around the junction.  For example, consider the two different initial conditions shown in Fig.~\ref{fig:spacer}a: one is the abrupt junction simulated in our numerics, where a single bond is restored at $t=0$; the other contains a ``spacer'' of $n$ bonds, all of which are turned on at $t=0$.  The total energy current integrated over bonds between these two reservoirs must be exactly the same in these two cases, up to some long time (we assume that the reservoir size $L_0 \gg n$.)  But for a long spacer $n \gg 1$, the spatial profile of the current will look different at short times than for $n=1$: for the long spacer, the current consists of one right-moving patch from the left lead and one left-moving patch from the right lead, as shown schematically in Fig.~\ref{fig:spacer}b.  For $n=1$, the current is expected to be spatially monotonic, or at least not to decouple into these separated regions.

This argument indicates that the nontrivial time evolution of spatially integrated current is exactly the same whether the right-movers and left-movers pass through each other or are spatially separated, i.e., treating the problem as two separate pulses and adding them together gives exactly the right total current.  We now give a (non-rigorous) picture for how a steady state described by a Stefan-Boltzmann function can exist independent of the ``pulse shape'', i.e., the spatial distribution of energy current, which is certainly sensitive to interactions.  Consider each large reservoir, now of size $L$, as made up of many contiguous sub-reservoirs of size $L_0$ (Fig.~\ref{fig:spacer}b).  We can pair sub-reservoirs as shown so that the total problem is a combination of copies translated by $L_0$.  The point of the small sub-reservoirs is that the observation time for the steady state is long enough that the sub-reservoirs are in the long-time limit where $F$ is defined, although the total reservoir is not.  The shape of these contributions could be modified by being in a non-trivial background, but their total weight is not, and as we are at long times, we expect the pulse shape to be broad compared to $L_0$.  Now whatever the shape of the contribution from a pair of sub-reservoirs is, its total weight is $L_0 (f(T_L) - f(T_R))$.  When we add together many translated copies of the same shape, we obtain a constant, since by the Poisson summation formula
\begin{equation}
\sum_{n=-\infty}^\infty g(x+n a) \approx C = {\int_{-\infty}^\infty g(x)\,dx \over a}
\end{equation}
for functions $g$ that do not vary strongly on the scale of $a$.  So this constant is exactly $j_E^\infty = f(T_L) - f(T_R)$, the desired result. The conformal limit is a case where the subreservoir size can be taken to be arbitrarily small.  In words, the steady state exists in a time window where all that matters is the total energy per length of right-movers emitted from the left, less that of left-movers emitted from the right.

\begin{figure}[t]
\includegraphics[width=0.95\linewidth]{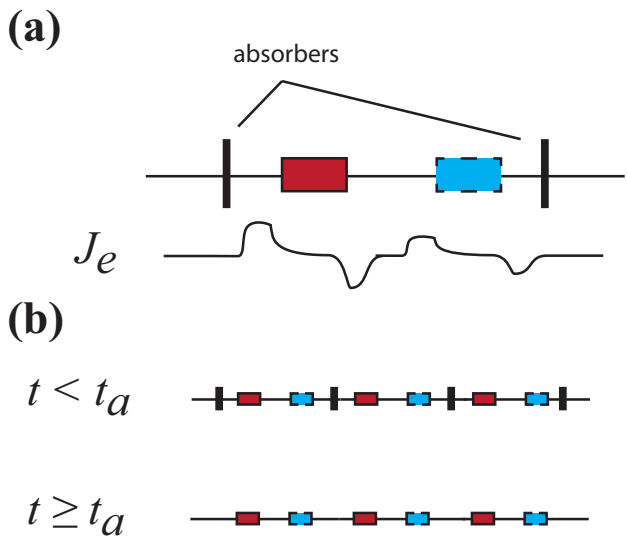}
\caption{Protocol to prepare a translation-invariant final state with energy current given by the Stefan-Boltzmann prediction. (a) Consider one hot lead and one cold lead, each of which radiates both right-movers and left-movers.  To create a state that mimicks the original two-lead geometry, place non-unitary ``aborbers'' as shown.  (b) Replicate the geometry of (a) in both directions.  The absorbers are maintained in place long enough (until some time $t_a$) to absorb half of the excitations from each initial reservoir; which half are absorbed (right-movers or left-movers) depends on the reservoir's temperature.  The absorbers are then removed and further evolution is unitary.  Energy current is conserved and, if the dispersion of velocities leads to a translation-invariant final state, that final state must have energy current given by the Stefan-Boltzmann prediction.}
\label{fig:protocol} 
\end{figure}

We can give a specific example for which this picture is correct by constructing a geometry, different from the original geometry of two semi-infinite leads, for which the time evolution can be shown to give a steady-state described by the Stefan-Boltzmann relation, even in the presence of interactions.  This sidesteps the difficulty of solving the time evolution from the non-translation-invariant initial state--it is difficult to prove in general even that a translation-invariant steady state exists.  We would like to make a translation-invariant system of interacting particles whose final state has left-movers originating at a different temperature from right-movers.  To do so, consider the process in Fig.~\ref{fig:protocol}a.  Suppose for simplicity that the model has a finite range of particle velocities and that there is a ``vacuum'' state with no particles.  Two regions of different temperatures are spaced by a large enough region of vacuum that the left-movers from the right region fail to interact with the left-movers from the other until both types of particles have moved out of their original reservoir regions.  Particle absorbers, which could be extra lengths of spin chain for example, are placed at the left and right ends of the system and absorb left-movers from the left region and right-movers from the left region.

After this absorption, the absorbers are removed and the remaining evolution is unitary.  Repeat this arrangement in a discretely translation-invariant way, as in Fig.~\ref{fig:protocol}b.  The further time evolution will preserve the total energy current, even though the particles coming from different reservoirs will now interact.  Assuming that a homogeneous energy current is reached in the final state because of the dispersion of particle velocities, the energy current in this final state has to be $c \left[ f(T_1) - f(T_2) \right]$, where $c$ is the volume fraction occupied by the reservoirs and $f$ is the directed radiated energy current per unit length.  In other words, the conservation of energy current means that, in this specific example where left-movers and right-movers are drawn from different temperatures, the final state is described by a Stefan-Boltzmann function.  However, the existence of the same energy current in the actual two-reservoir geometry is so far more difficult to establish.


While particle velocities are certainly modified by the density of particles from the other reservoir in an interacting system, this modification need not alter the total energy current at the boundary, as in the example above.  A quantum field theory approach to thermal steady states leads in some cases to a factorization of the density matrix from which the existence of a Stefan-Boltzmann function follows, although integrability and conserved energy current seem to play less of a role in this approach~\cite{doyonhouches}.  It should be noted that applying this field-theory approach to the steady-state energy current in the massive sine-Gordon model gives behavior that slightly violates the existence of a Stefan-Boltzmann function~\cite{doyonviolate}, at the level of a few percent.  It is not clear whether this discrepancy indicates a fundamental difference in the type of steady state (as, analytically, solving the time evolution for all times to see the unique steady state emerge is not possible in either approach).  Finally, we note that the same logic presented in this section would apply to charge currents in a homogeneous system with a conserved charge current, but the XXZ model's charge current is not conserved (does not commute with the Hamiltonian), although there is a Drude weight.

\section{Bethe ansatz approximation for the steady-state current}
\label{sec:bethe}
Within the accuracy of our numerics, the steady state current does not depend jointly on $T_L$ and $T_R$ but rather is just the difference of the function $f$ evaluated at the two temperatures.  It is known that at low $T$ this property should hold as a result of conformal invariance,\cite{doyon} with $f \sim T^2$.  Earlier in this paper we gave a somewhat lengthy analytic argument suggesting this result; in order to make it less mysterious, we now show that a relatively simple approximation for the isotropic XXX ferromagnet ($\Delta=-1$) gives a reasonable description of $f(T)$ at all temperatures and is exact in the low-temperature limit. 

\begin{figure}[t]
\includegraphics[width=0.95\linewidth]{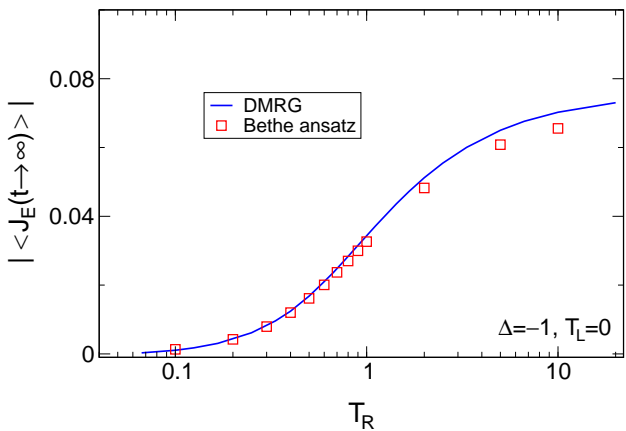}
\caption{Comparison for the isotropic XXX ferromagnet ($\Delta=-1$) of two estimates for the function $f$ whose differences give the steady-state energy current $J_E$ via $J_E = f(T_L) - f(T_R)$ between reservoirs $T_L$ and $T_R$.  One estimate is obtained from the DMRG results, using reference temperatures .  The other is obtained from the numerical solution of the thermodynamic Bethe ansatz equations and the formula Eq.~(\ref{eq:tbaresult}).}
\label{fig:bethe} 
\end{figure}

For a spatially uniform gas of free particles with a particle distribution function $\rho(v)$ in velocity space, the Stefan-Boltzmann function $f_\tn{SB}$ can be obtained as
\begin{equation}
f_\tn{SB}(T) = \int_0^\infty \rho(v) v E(v)\,dv~,
\label{eq:f_free}
\end{equation}
where $E(v)$ is the energy of a particle with velocity $v$ and the integral is over only right-moving particles.  This has the units of an energy current (an energy per time) if $\rho$ has units of particle number per length per velocity.

Eq.~(\ref{eq:f_free}) motivates the following approximation: the thermodynamic Bethe ansatz solution of the XXX model by Takahashi\cite{takahashi_tba,takahashi2} is in terms of particles named ``magnons'' and ``strings''.  In general, these particles are not independent because of the Bethe conditions on phase shifts.  However, at low temperature and for ferromagnetic interactions, the thermodynamic state is a dilute gas of magnons and strings, and we might hope that a formula similar to Eq.~(\ref{eq:f_free}), generalized to multiple kinds of particles, is a good starting point.

\begin{figure}[t]
\includegraphics[width=0.95\linewidth,clip]{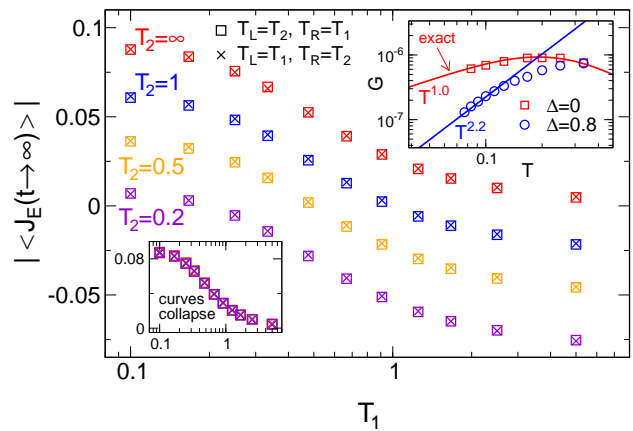}
\caption{ Steady state current for $\lambda=1$, $b=0$ but different bulk interactions $\Delta_L=0.3, \Delta_R=0.8$ and different couplings $J_L=1, J_R=0.808$ to tune the backscattering at the interface to zero.\cite{backscattering} The magnitude of the current is the same for $T_L=T_1, T_R=T_2$ and $T_L=T_2, T_R=T_1$. The curves at different $T_2$ collapse (left Inset), indicating that Eq.~(\ref{eq:f}) still holds. Right Inset: Linear thermal conductance of two identical XXZ chains connected by a strong barrier $J_0=0.002$. The low-$T$ behavior is consistent with $T^{2/K-1}$ where $K$ is the Luttinger liquid parameter.\cite{kanefisher} At $\Delta=0$ we repoduce the exact result of Ref.~\onlinecite{bruneau} for any $T$. }
\label{fig:inhom}
\end{figure}

In the notation of Refs.~\onlinecite{takahashi_tba,takahashi2}, which introduces a variable $x$ that parametrizes the momentum $k$ via
\begin{equation}
e^{i k} = {x+\frac{i}{x} - i}~,
\end{equation}
the needed quantity is
\begin{equation}
f(T) = \int_{x>0} \rho_n(x) v_n(x) E_n(x)~,
\label{eq:tbaresult}
\end{equation}
where $n$ ranges over the different types of excitations (magnons and strings)  and $\rho, v, E$ are the density, velocity, and energy.  Note that there is an ambiguous additive constant in $f$.  We evaluate only for right-moving excitations based on a Landauer-type picture where the steady-state can be viewed as the combination of right-movers from the left lead and left-movers from the right lead.  The velocity $v_n$ was obtained as $dE_n/dk$ and the energy for an excitation of total momentum $K$ is
\begin{equation}
E_n = {2 \over n} (1 - \cos K)~.
\end{equation}

We have solved the standard TBA equations for $\rho_n$ using the numerical method introduced of Schlottman.\cite{schlottmann} The result of evaluating this form for $f$ is surprisingly good for the ferromagnetic case: it is correct at low temperature and underestimates the correct value by about 10\% at high temperature (see Fig.~\ref{fig:bethe}). For the XXX antiferromagnet, the result is much worse and fails to reproduce the CFT result at low temperature, which is natural as the low-temperature state is now not dilute in terms of these particles. However, the agreement is improved at low temperature by using a group velocity derived from the dressed excitation energy.  A more complete comparison to this approach, including other values of the anisotropy parameter, is currently underway.

A full explanation of our numerical results from the Bethe ansatz is an open problem, and just taking the right-movers (as done here) does not satisfy the Bethe conditions on the phase shifts of the particles. Nevertheless, the quantitative agreement between the estimate from Eq.~(\ref{eq:tbaresult}) and the DMRG result means that the steady-state current is close in the ferromagnet to a free-particle interpretation although the particles and their densities are rather complicated.

\section{Inhomogeneous systems}
\label{sec:inhomogeneous}
We finally investigate systems which are not translationally invariant.  Tunneling across a barrier between Luttinger liquids has been well studied by bosonization and other field-theoretic methods in the low-energy limit, and beyond just verifying these predictions, numerics allow a determination of when the asymptotic properties accessible by bosonization become apparent.  We start by studying the effects of different bulk parameters $J_{n<0}=J_L$, $\Delta_{n<0}=\Delta_L$, $J_{n\geq 0}=J_R$, $\Delta_{n\geq0}=\Delta_R$ (and $b_n=0$). If these parameters are chosen such that the renormalized Fermi velocities in the left and right halves coincide, backscattering due to the barrier (which is naturally present at the interface $n=0$ \cite{katharina}) can be tuned to zero \cite{backscattering}, and Luttinger liquid (LL) boundary physics \cite{kanefisher} is absent. Results are displayed in Figure \ref{fig:inhom}, indicating that Eq.~(\ref{eq:f}) still holds.

If two homogeneous XXZ chains $J_{n\neq0}=1, \Delta_n=\Delta$ are connected through a barrier $J_0$, the linear thermal conductance $G = \partial_{T_L}\langle J_E(t\to\infty)\rangle|_{T_L=T_R=T}$ is expected to feature a low-$T$ power law $T^{2/K-1}$ with $K$ being the LL parameter.\cite{kanefisher} For $\Delta=0$, Eq.~(\ref{eq:f}) still holds even if $J_0<1$,\cite{bruneau} and $2/K-1=1$ just reflects the asymptotic $T^2$ behavior of $f$. Our data for $G$ in presence of interactions is consistent with $T^{2/K-1}$. This is illustrated in the Inset to Figure \ref{fig:inhom} where $J_0$ is chosen small so that the scale on which LL boundary effects manifests becomes large.\cite{llscale}
Eq.~(\ref{eq:f}) still holds approximately above this scale.  There are obviously many possible bulk and tunneling parameters that can be studied, and we reserve a comprehensive study for future work; the main point is to note that the Luttinger liquid tunneling physics and other subtle properties can be accessed via our approach.

\section{Summary}
In this paper we provided evidence for a connection between nonequilibrium and linear-response thermal transport properties of isolated infinite spin chains (or equivalently, interacting spinless fermions): (1) The energy current of a system which initially features a temperature gradient $T_L\neq T_R$ saturates to a finite value if the equilibrium thermal Drude weight $D$ is finite, and (2) The value of the steady-state current at arbitrary $T_{L,R}$ is of the functional form $J_E=f(T_L)-f(T_R)$, i.e.~it is completely determined by the linear thermal conductance. This can be viewed as a generalized Stefan-Boltzmann law describing freely moving quasiparticles; for the XXX ferromagnet, $f$ can be computed via thermodynamic Bethe ansatz in good agreement with the numerics. Our data suggests that $D>0$ for a nonintegrable dimerized chain (or that the current correlation function decays on a hidden large temperature-independent time scale).

\emph{Acknowledgments} --- We are indebted to E.~Altman, B.~Doyon, F.~Essler, F.~Heidrich-Meisner, V.~Meden, K.~Sch\"onhammer, and D.~Schuricht for fruitful discussions and comments and acknowledge support by the Deutsche Forschungsgemeinschaft via KA3360-1/1 (C.K.) as well as by the AFOSR MURI on ``Control of Thermal and Electrical Transport'' (R.I.), the Nanostructured Thermoelectrics program of LBNL (J.E.M. and C.K.), and the Simons Foundation (J.E.M.).

\appendix
\section{Steady state for a Maxwellian distribution}

We would like to understand in a simple example how the steady-state current arises outside the conformal limit, i.e., when particle velocities are variable.  Consider a system of classical non-interacting particles that at time $t=0$ has the Maxwellian distribution
\begin{equation}
f(x,v,0) = \begin{cases}
c \exp(-\alpha v^2/2)\ {\rm if}\ x \in [-L/2,L/2], \\
0\ {\rm otherwise}.
\end{cases}
\end{equation}
The Boltzmann equation contains only the streaming term, with the result that the function $f(x,v,t)$ is simply equal to $f(x-vt,v,0)$: the number of particles with a given velocity $v$ at the point $x$ and time $t$ is given by the number of particles with that velocity at spatial point $x-vt$ at time 0.

We would like to compute the energy current
\begin{equation}
j_E(x,t) \equiv \int_0^\infty\,dv\,v \left( {v^2 \over 2} \right) f(x,v,t).
\end{equation}
The integral over the initial distribution will contribute if $x-vt \in [-L/2,L/2]$, which means
\begin{equation}
v \in [{-L/2 + x \over t}, {L/2 + x \over t}].
\end{equation}
Assuming $x \geq L/2$, the energy current is
\begin{eqnarray}
j_E(x,t) &=& - \int_{(-L/2 + x) / t}^{(L/2 + x) / t}\,dv\,{v^3 \over 2} c e^{- \alpha v^2 / 2} \\
&=& -c {2 + \alpha v^2 \over \alpha^2} e^{-\alpha v^2 \over 2} \Big ]_{(-L/2 + x) / t}^{(L/2 + x) / t}.
\label{energycurrent}
\end{eqnarray}

The Stefan-Boltzmann function $f$ as defined above is given by the total right-moving energy current per length, or
\begin{equation}
f = \int_0^\infty\,dv\,{v^3 \over 2} c e^{- \alpha v^2 / 2} = {2 c \over \alpha^2}.
\end{equation}
Now the question is whether (\ref{energycurrent}), evaluated at the right edge of the reservoir, is equal to $f$ for some period.
We have
\begin{equation}
j_E(L/2,t) = c \left[ {2 \over \alpha^2} - {2 + \alpha (L^2/t^2) \over \alpha^2} e^{-\alpha L^2 \over 2 t^2} \right].
\end{equation}
We see that this is indeed equal to $f$ while
\begin{equation}
t \ll  L \sqrt {\alpha \over 2},
\end{equation}
and that the steady state described by $f$ persists forever if $L \rightarrow \infty$.


\begin{thebibliography}{*}
\bibitem{levelstat}
    This is supported by level statistics.\cite{integrabilitypaper}
\end{thebibliography}
\end{document}